\documentclass[aps,prd,preprint,superscriptaddress,showpacs]{revtex4}
\usepackage{epsf,epsfig,graphics,graphicx,pdfpages}
\usepackage{verbatim,color,ulem}
\newcommand{\be}{\begin{equation}}
\newcommand{\ee}{\end{equation}}
\newcommand{\bea}{\begin{eqnarray}}
\newcommand{\eea}{\end{eqnarray}}
\newcommand{\ba}{\begin{array}}
\newcommand{\ea}{\end{array}}
\input epsf
\usepackage{amsmath,amssymb}
\begin{document}

\title{Holographic influence functional and  its application to decoherence  induced by quantum critical theories}
\author{Chen-Pin Yeh}
\email{chenpinyeh@mail.ndhu.edu.tw}
\affiliation{Department of
Physics, National Dong-Hwa University, Hualien, Taiwan, R.O.C.}

\author{Jen-Tsung Hsiang}
\email{cosmology@gmail.com}
\affiliation{Department of Physics,
National Dong-Hwa University, Hualien, Taiwan, R.O.C.}
\affiliation{Center for Theoretical Physics, Fudan University,
Shanghai, China}

\author{Da-Shin Lee}
\email{dslee@mail.ndhu.edu.tw}
\affiliation{Department of Physics,
National Dong-Hwa University, Hualien, Taiwan, R.O.C.}

\begin{abstract}
The dynamics of a particle influenced by strongly-coupled quantum
critical theories is studied by the holographic approach. A
real-time prescription for the AdS/CFT correspondence in the
context of nonequilibrium physics is proposed from mainly the
field theoretic consideration, and the associated holographic
influence functional is obtained. We then study the decoherence
dynamics of a particle induced by the quantum critical theories
with a dynamical exponent $z$. We find that as $z$ increases, the
decoherence effect becomes less significant. The similar behavior
is found as we heat up the environment. However, in this case the
decoherence effect is enhanced not only by its strong coupling
constant but also from the finite temperature effect. Finally a
comparison is made with the result of a particle influenced by the
free field.

\end{abstract}


\maketitle
\section{Introduction}
Originally proposed as a holographic duality between the string
theory and quantum field theory\cite{AdSCFT}, AdS/CFT is found
inspiring in understanding strongly coupled field theory. A well
established dictionary is the relation between classical gravity
action and the generating function for large $N$ quantum field
theory (QFT) in the large coupling regime. This gives us a
prescription to calculate the correlators of strongly coupled QFT
by evaluating the corresponding classical gravity action. In
particular, AdS/CFT correspondence has been applied to understand
the strong coupling problems in condensed matter systems and the
hydrodynamics(see~\cite{Hartnoll_09,Rangmanai_09} for reivews).
Independently in \cite{Herzog:2006gh,Gubser_06,Teaney_06}, it is
also applied to explore the dissipation behavior of a particle
moving in a strongly coupled environment. In these studies, the
end point of the string on the boundary of the AdS black hole
serves as a probe particle. Then more works are drawn to realizing
the fluctuations of this end point of the string as Brownian
motion in more general
backgrounds~\cite{Son:2009vu,gubserqhat,ctqhat,Caceres:2010rm,Giecold:2009cg,CasalderreySolana:2009rm,Atmaja:2010uu,Das:2010yw,Gursoy:2010aa,Ebrahim:2010ra,CaronHuot:2011dr,Kiritsis:2011bw,
Fischler:2012ff,Tong_12,Edalati:2012tc,Atmaja:2012jg,Sadeghi:2013lka,Atmaja:2013gxa,Banerjee:2013rca,Giataganas:2013hwa,mirror,
Kiritsis:2013iba,Chakrabortty:2013kra,Sadeghi:2013jja,Giataganas:2013zaa,Sadeghi:2014lha,Fischler_14}.
For a general review on application in the non-equilibrium
dynamics, please see \cite{Holographic QBM}.

From the aspect of the open systems in the field theory, the
effects of environmental degrees of freedom on the particle can be
accounted for by the method of Feynman-Vernon influence
functional~\cite{Fv}. In this approach the environment variables
are integrated out from the full density matrix to derive the
evolution of the reduced density matrix with their effects encoded
in the influence functional~\cite{Leggett,GSI}. This tracing-out
process can be carried out by the path integral within the
closed-time-path formalism (the so-called Schwinger-Keldysh
formalism)~\cite{SC}. However, usually the influence functional
can at best be perturbatively calculated for a weakly coupled
environment. The idea of the paper is to employ the holography
method to find the influence functional of the strongly coupled
environment.

We consider a probed particle as the system in a strongly-coupled
quantum critical theory  with general dynamical scaling $z$, in
which the theory is invariant under the scaling,
  \be
  t \rightarrow \mu^z t,~~x\rightarrow \mu x  \, .
  \ee
At zero temperature, the holographic description consists of a
probed string moving in the classical Lifshitz geometry (see
\cite{mirror} and references therein) with the metric,
   \be
ds^2=-r^{2z}dt^2+\frac{dr^2}{r^2}+r^2d\vec{x}^2 \, .
   \ee
The holographic closed-time-path formalism is developed by Herzog
and Son in \cite{Son_09} and later by Skenderis and Ree in
\cite{Ree_08}. We here extend the approach in \cite{Son_09} to the
Lifshitz black hole background, but the correlators are obtained
from the field-theoretic considerations. This prescription may
provide a more transparent bridge in the AdS/CFT correspondence
within the context of nonequilibrium field theory.

As an application, we use the holographic influence functional
to study the quantum decoherence in a strongly coupled environment.
Quantum decoherence is an ubiquitous phenomenon owing to an
unavoidable interaction between the system of interest and the
environment that has a huge number of the degrees of freedom.
Recently many efforts have been devoted to the experimental
realization of quantum computers, in which the central obstacle is
to prevent the degradation of the quantum coherence of the computer
by the disturbance from the environment~\cite{NI}. This environment-induced effect has been studied in
the framework of quantum open systems~\cite{ZU,BAR,HS,Lee_06}. For
example, the quantized electromagnetic fields can result in the
stochastic motion of the classical charged
particle~\cite{Lee_08,Lee_12,Lee_09}. Additionally they give rise to
decoherence effects on the wavefunction of the quantum charged
particle~\cite{BAR,HS,Lee_06}. The effects can be observed in the
interference pattern through the phase shift and the contrast change
of the fringes as the consequence of quantum fluctuations of
electromagnetic fields. One of the main focuses of this paper is to
study how the interference of the particle states are affected by
strongly coupled environment in its vacuum and thermal states.

Our presentation is organized as follows. In next section, we
introduce the closed-time-path formalism. By tracing out the
environment degrees of freedom in the full density matrix of the
system and environment, we obtain the reduced density matrix of
the system. In the linear response region, the environment effects
are encoded in the associated influence functional. The
holographic influence functional is then constructed with the
prescription on the dual theory of a probed string in the
classical Lifshitz geometry in Sec.~\ref{sec2}. In
Sec.~\ref{sec3}, we apply the obtained influence functional to
study quantum decoherence of a particle in quantum critical
theories. This reduction of the interference is measured by the
decoherence functional, see \cite{Lee_06}. Concluding remarks are
in Sec.~\ref{sec4}.

\section{closed-time-path  formalism} \label{sec1}
In the context of nonequilibrium physics, it is often convenient
to use the so-called closed-time-path formalism (Schwinger-Keldysh
formalism). Here we express the overall effects of the environment
by the path integral along a closed time contour in the complex
time plane, and then the expectation value for the system states,
in particular, the nonequilibrium correlators can be obtained
accordingly. To see it, let us consider that ${\hat \rho}(t)$ is
the density matrix of the system-environment, and then it evolves
unitarily according to
\begin{equation}
{\hat \rho} (t_f) = U(t_f, t_i) \, {\hat \rho} (t_i) \, U^{-1} (t_f,
t_i )
\end{equation}
with $ U(t_f,t_i) $ the time evolution operator of the whole system.
We then assume that the initial density matrix at time $t_i$ can be
factorized as
\begin{equation}\label{initialcond}
    \rho(t_i)=\rho_{q}(t_i)\otimes\rho_{{F}}(t_i)\,
    ,
\end{equation}
where $q$ and $F$ are system and environment variables
respectively. Initially the environment field $F$ is in thermal
equilibrium at temperature $T=1/\beta $ with the density matrix
$\rho_{{F}}(t_i)$ given by taking the form
\begin{equation}\label{initialcondphi}
    \rho_{F}(t_i)=e^{-\beta H_{F}}\,,
\end{equation}
where $H_{F}$ is the Hamiltonian for the environment field. The
initial vacuum state with zero temperature can be reached by
taking $T\rightarrow 0$ limit.

After we trace out the environmental degrees of freedom in
the full density matrix $\hat{\rho}$,  the  environment effect on
the system of interest is then realized in the reduced density matrix
$\rho_r$. The reduced density matrix $\rho_r$ is given by
\begin{eqnarray}
 \rho_r({q}_f,\tilde{{q}}_f,t_f)&=&\int\!d\tilde{{F}}\;\bigl<{q}_f,\tilde{{F}}\big|\rho(t_f)\big|\tilde{{q}}_f,\tilde{{F}}\bigr>\nonumber \\
                   &=&\int\!d\tilde{{F}}\int\!d{q}_1\,d{F}_{1}\int\!d{q}_2\,d{F}_{2}\;\bigl<{q},\tilde{F}\big|U(t_f,t_i)\big|{q}_1,F_1\bigr>\nonumber\\
                   &&\qquad\qquad\times\bigl<{q}_1,F_1\big|\rho(t_i)\big|{q}_2,F_{2}\bigr>\bigl<{q}_2,F_{2}\big|U^{-1}(t_f,t_i)\big|\tilde{{q}},\tilde{F}\bigr>\nonumber\\
                   &=&\int\!d{q}_1\,d{q}_2\int\!d\tilde{F}\,dF_{1}\,dF_{2}\int^{{{q}}_f}_{{q}_1}\!\mathcal{D}{q}^+\int^{\tilde{{q}}_f}_{{q}_2}\!{\cal D}{q}^-\int^{\tilde{F}}_{F_{1}}\!\mathcal{D}F^+\int^{\tilde{F}}_{F_{2}}\!\mathcal{D}F^-\nonumber\\
                   &&\qquad\qquad\times\int^{F{1}}_{F_{2}}\!\mathcal{D}F^{\beta}\;\exp\biggl[i\int_{t_i}^{t_f}dt\;L[{q}^+,F^+]-L[{q}^-,F^-]\biggr]\nonumber\\
                   &&\qquad\qquad\qquad\qquad\qquad\times\exp\biggl[i\int_{t_i}^{t_i-i\beta}dt\;L_F [F^{\beta}]\biggr]\;\rho_{q}({q}_1,{q}_2,t_i)\,.
\end{eqnarray}
Here we have inserted the complete set of eigenstates,
$\big|q_1,F_1 \bigr>$ and $\big|q_2,F_2 \bigr>$, which are given
by the direct product of the system and environment states. The
matrices of time evolution operators, $U(t_f,t_i)$ and
$U^{-1}(t_f,t_i)$ can be expressed by the path integral along the
forward and backward time evolution, represented by ${q}^+$,
$F^+$, and ${q}^-$, $F^-$, respectively. The density matrix for
the thermal state of the environment corresponds to the evolution
operator matrix of the field $F^{\beta}$ along a path between the
complex time $t_i$ and $t_i-i\beta$ (see Fig.~1). Thus, the
Green's functions of the field possess the periodicity as the
result of the cyclic property of the trace as well as the bosonic
nature of the field operator.
\begin{figure}
\centering
\includegraphics[width=\columnwidth]{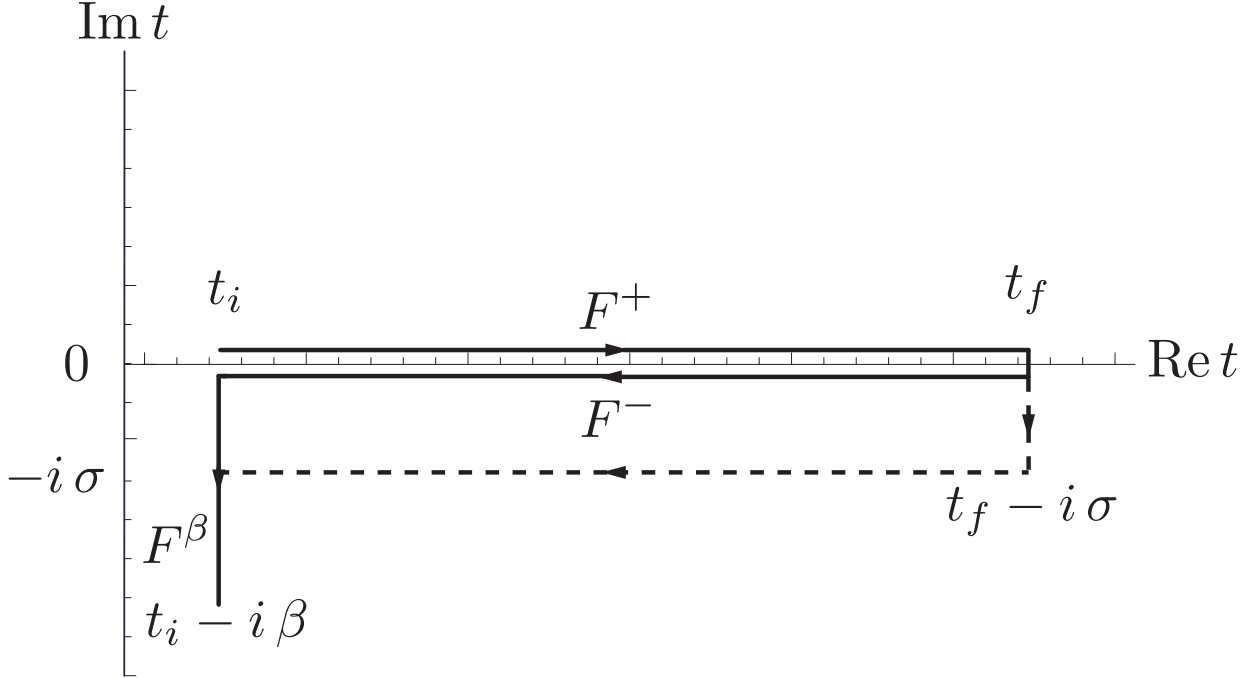}
\caption{{The contour ($C$) of the closed-time-path formalism (Solid
line). The variable field labeled by $+$ ($-$) lives on the upper
(lower) time axis, while the field labeled by $\beta$ lives along
the imaginary time axis. The contour can be generalized to the
finite $\sigma$ depicted in (Dotted line). }}
\end{figure}

We consider the system linearly coupled to an environment field
$F$. The full Lagrangian is
  \be
   L(q,F)=L_q[q]+L_F[F]+qF\, .
  \ee
We write the evolution of the reduced density matrix in the form
\begin{equation}
\rho_r({q}_f,\tilde{{q}}_f,t_f)=\int\!d{q}_1\,d{q}_2\;\mathcal{J}({q}_f,\tilde{{q}}_f,t_f;{q}_1,{q}_2,t_i)\,\rho_{q}({q}_1,{q}_2,t_i)\,,\label{evolveelectron}
\end{equation}
where the propagating function
$\mathcal{J}({q}_f,\tilde{{q}}_f,t_f;{q}_1,{q}_2,t_i)$ is
\begin{equation}\label{propagator}
    \mathcal{J}({q}_f,\tilde{{q}}_f,t_f;{q}_1,{q}_2,t_i)=\int^{{q}_f}_{{q}_1}\!\!\mathcal{D}{q}^+\!\!\int^{\tilde{{q}}_f}_{{q}_2}\!\!\mathcal{D}{q}^-\;\exp\left[i\int_{t_i}^{t_f}dt\left(L_{q}[{q}^+]-L_{q}[{q}^-]\right)\right]\mathcal{F}[{q}^+,{q}^-]\,,
\end{equation}
and the influence functional $\mathcal{F}[{q}^+,{q}^-]$ is defined
as
\begin{eqnarray}
\mathcal{F}[{q}^+,{q}^-]   &=& \mathcal{N}\int\!d\tilde{F}\,dF_{1}\,dF_{2} \int^{\tilde{F}}_{F_{1}}\!\mathcal{D}F^+\int^{\tilde{F}}_{F_{2}}\!\mathcal{D}F^-\int^{F{1}}_{F_{2}}\!\mathcal{D}F^{\beta}\;\exp\biggl\{i\int_{t_i}^{t_i-i\beta}dt\;L_F[F^{\beta}]\biggr\}\nonumber\\
                   &&\qquad\qquad\qquad\times\exp\biggl\{i\int_{t_i}^{t_f} dt\; L_F[F^{+}]+ {q}^+  F^+ ( t ) - L_F[F^{-}]-{q}^-  F^- ( t )\biggr\}
                   \nonumber\\
                    &=& \int\!d\tilde{F}\, \int_C
                    \!\mathcal{D}F^C \exp
                    \left[i\int_C d t \,  L_F[F^{C}]+ {q}^C  F^C
                    \right]/  \int\!d\tilde{F}\, \int
                    \!\mathcal{D}F^{\beta} \, \exp\left[
                    i\int^{t_i-i\beta}_{t_i}  d t \,  L_F[F^{\beta}]\right]
                    \, .\nonumber\\  \label{IF}
\end{eqnarray}
In the end the influence functional can be expressed as a path
integral along the contour $C$, as shown in Fig.~1, with the
periodical boundary condition $F^C (t_i)=F^C(t_i-i
\beta)=\tilde{F}$. The variables $F^{+}$, $q^{+}$ ($F^{-}$,
$q^{-}$) are associated with the upper (lower) branch of the
closed-time contour while $F^{\beta}$ lives along the imaginary
time axis. The normalization $\mathcal{N}$ has been chosen so that
$\mathcal{F}[0,0] =1$. After carrying out the path integral over
the environmental degrees of freedom, we find that
 \begin{equation}
  \label{inf}
  \mathcal{F}[{q}^+,{q}^-]   = \exp\biggl\{
-\frac{1}{2}\int_{C \, {\rm on}\, \pm} dt \!\!\int_{C\, {\rm on}\,
\pm} \!dt'\; \Bigl[
{q}^C(t)\,\bigl<F^{C}(t)F^{C}(t')\bigr>\,{q}^C(t')\Bigr]
\biggr\}\,,
 \end{equation}
where we have only kept the terms to the quadratic order of the
system variable within the liner response region. The
contour-ordered Green's functions are defined as
\begin{equation}
\bigl< F^{C}(t)F^{C}(t')\bigr>= \bigl<F(t)F(t')\bigr> \, \theta_C
(t-t')+ \bigl<F(t')F(t)\bigr>\, \theta_C (t'-t) \,,
\end{equation}
where the  $\theta_C$ function  is defined according to the path
ordering along the contour.  The correlators are defined with
respect to the thermal state of the environment field.  The
periodicity boundary condition in the path integral along the
contour is a consequence of the thermal bath of the bosonic field
\begin{equation}\label{E:dkejr}
\bigl< F^{C}(t_i)F^{C}(t')\bigr>= \bigl< F^{C}(t_i-i
\beta)F^{C}(t')\bigr>
\end{equation}
leading to
 \begin{equation}
 \label{periodicity}
 \bigl<F(t-i\beta)F(t')\bigr>= \bigl<F(t')F(t)\bigr> \, ,
 \end{equation}
which obeys the Kubo-Martin-Schwinger condition. Expressed in
terms of times at $\pm$ branches, the influence functional
$\mathcal{F}\left[{q}^{+},{q}^{-}\right]$
 can be written in terms of real-time Green's functions,
\begin{align}\label{influencefun2}
{\mathcal{F}}\left[{q}^{+},{q}^{-}\right]&= \exp\biggl\{
-\frac{i}{2}\int_{t_i}^{t_f} dt\!\!\int_{t_i}^{t_f} \!dt' \Bigl[
{q}^+(t)\,G^{++}(t,t') \,{q}^+(t')\Bigr.\biggr.
-{q}^+(t)\,G^{+-}(t,t')\,{q}^-(t') \nonumber\\
&\qquad\qquad \qquad\qquad \qquad-
{q}^-(t)\,G^{-+}(t,t')\,{q}^+(t')
\biggl.\Bigl.+{q}^-(t)\,G^{--}(t,t')\,{q}^-(t')\Bigr]\biggr\}\,.
\end{align}
The various correlation functions are defined as \bea
\label{correlator}
    && i\,G^{+-}(t,t')=\langle F(t')F(t)\rangle \, , \nonumber\\
    && i\,G^{-+}(t,t')=\langle F(t)F(t')\rangle \, ,\nonumber\\
    &&i\,G^{++}(t,t')=\langle F(t)F(t')\rangle\theta(t-t')+\langle
    F(t')F(t)\rangle\theta(t'-t)\, , \nonumber\\
    && i\,G^{--}(t,t')=\langle F(t')F(t)\rangle\theta(t-t')+\langle
    F(t)F(t')\rangle\theta(t'-t) \, .
    \eea
The path integral expression of the influence functional
in~(\ref{IF}) can be understood as the field
propagator of $F^{\pm}$ with an source $q^{\pm}$ in their respective
branches. However the field evolves unitarily, giving
$\mathcal{F}\left[{q},{q}\right]=1$ due to the fact that the real
time propagators on the $\pm$ branches cancel each other when
setting $q^\pm=q$. Thus, the unitarity property of the field
evolution gives a relation among various Green's functions.
\begin{equation}
G^{++}(t,t') +  G^{--}(t,t')- G^{+-}(t,t')- G^{-+}(t,t') =0
\, . \label{ur-identity}
\end{equation}
This relation  can be checked straightforwardly
from~(\ref{correlator}) in the real-time expressions. For the
time-translation invariant states,  the Fourier transform of the
Green's function is defined as $G(\omega)=\int dt\;
G(t,0)e^{i\omega t}$, where the  identity in \eqref{ur-identity}
becomes
\begin{equation}
\qquad\qquad\qquad G^{++}(\omega) +  G^{--}(\omega)-
G^{+-}(\omega)- G^{-+}(\omega) =0 \, . \label{u-identity}
\end{equation}
We will see later that the periodicity
condition~(\ref{periodicity}) and the unitarity
property~(\ref{u-identity}) of a bosonic thermal bath are
fundamental requirements to construct the holographic influence
functional.

\section{Holographic influence functional}\label{sec2}
In this section, we will employ the holographic method to derive
the influence functional for the Brownian particle. The dual
description of the particle coupled to quantum critical
environment at finite temperature is the string propagating in the
Lifshitz black hole. The idea \cite{Son_09} is to maximally extend
the black hole geometry in the Kruskal coordinates, which has two
asymptotic boundaries. By choosing appropriate boundary conditions
for the perturbations of the string $Q(t,r)$ in this background
geometry \cite{Son_09}, the classical on-shell action of the
string can be identified as the influence functional for the
Brownian particle in the boundary theory
   \be
   \label{gravity action}
   \mathcal{F} [q^+,
   q^-]=S_{gravity}\left(Q^+(t,r_b),Q^-(t,r_b)\right)\, .
   \ee
The variable $q (t)$ is the position of Brownian particle that
corresponds to the end point of the string perturbation $Q (t,r)$,
and $q^{\pm}(t)=Q^{\pm}(t,r_{b})$. The parameter $r_b$ is a cutoff
in the radial direction to render the action finite, and thus
represents the location of the boundary. In accordance with the
above closed-time-path formalism, $Q^+(t,r_1)$ and $Q^-(t,r_2)$
live in two respective regions with the different boundaries in
the maximally extended black hole geometry. These two regions will
be introduced later as shown in Fig.~2.

In the following, we will first introduce the maximally extended
Liftshitz black hole and then solve the classical equation of
motion for a probed string in this extended background. Instead of
imposing the infalling boundary conditions at the horizons as in
\cite{Son_09}, we will propose an alternative approach, which
makes use of the unitarity and periodicity properties of the
bosonic thermal field theory, to construct the influence
functional in (\ref{gravity action}). Thus we call it the
``semi-holographic" approach. The holographic influence functional
in \cite{Son_09} is rederived in the Appendix \ref{appen2} for
comparison. However this ``semi-holographic" approach provides a
more transparent method to set up the correspondence between
classical gravity theory and nonequilibrium strongly coupled
quantum field theory. Along the same line of thoughts we expect
that this approach will be straightforwardly used to obtain the
holographic influence functional for more general initial states
such as coherent or squeezed states \cite{squeeze}, and also for
the environmental degree of freedom that obeys the Fermi-Dirac
statistics.

\subsection{Maximally extended Liftshitz black hole}

We consider the Brownian particle coupled to quantum critical
theories at finite temperature \cite{Tong_12,mirror} in
3+1-dimension. The dual description is the string moving in the
4+1-dimensional Lifshitz black hole geometry with the metric
  \be
  \label{lifshitz bh}
  ds^2=-r^{2z}f(r)dt^2+\frac{dr^2}{f(r)r^2}+r^2d\vec{x}^2 \, ,
  \ee
where $f(r)\rightarrow1$ for $r\rightarrow\infty$ and $f(r)\simeq
c(r-r_h)$ near the black hole horizon $r_h$ with
$c=({z+3})/{r_h}$. The detailed form of $f(r)$ is not relevant
when we consider the low-frequency behavior of the perturbations.
The corresponding temperature of the back hole and also of the
boundary theory is
 \be
 \label{BHT}
\frac1T=\frac{4\pi}{z+3}\frac1{r_h^z} \, .
  \ee
The Kruskal coordinates are used to describe the maximally
extended geometry. It is convenient to first introduce a new
radial coordinate $r_{*}$
  \be
  r_*=\int dr\frac1{r^{z+1}f(r)} \, .
  \ee
Notice that when $z>0$, we have $r_*\rightarrow -\infty$ as
$r\rightarrow r_h$, and $r_*$ approaches a finite value
$r_*^{\infty}$ as $r\rightarrow \infty$. The metric \eqref{lifshitz bh} then becomes
  \be
   ds^2=r^{2z}f(r)(-dt^2+dr_*^2)+r^2d\vec{x}^2 \, .
  \ee
The Kruskal coordinates $(U,V)$ are defined as
   \be
   \label{Kruskal}
   V=e^{2\pi T(t+r_*-r_*^{\infty})}\, ,\,\,\, U=-e^{-2\pi T(t-r_*+r_*^{\infty})} \,
   .
   \ee
The product, $UV=-e^{4\pi T(r_*-r_*^{\infty})}$, depends only on
the radial coordinate $r$. The exterior of the black hole is the
region $V\ge 0$, $U\le 0$, where $UV=-1$ as $r\rightarrow\infty$
and $UV=0$ as $r=r_h$. This is called region $I$ (see Fig.~2 for
details). Notice that near the horizon $r_h$:
    \be
    r_*\simeq \frac{r_h^{-z}}{z+3}\ln(\frac{r}{r_h}-1) \, .
    \ee
Thus $UV\propto \frac{r}{r_h}-1$ and it changes sign as one moves
from the black hole exterior to the interior. We can analytically
continue the coordinates to either $U>0, V>0$ or $U<0, V<0$ which
corresponds to the black hole interior. However here we are
interested in another continuation that also describes the black
hole exterior with $U>0, V<0$, called region $I\!\!I$. Both region
$I$ and $I\!\!I$ contain the time-like infinity at
$r\rightarrow\infty$, where the boundary theory can be defined. We
label the spatial coordinate in  region $I$ as $({\vec x}_1,r_1)$
and in region $I\!\!I$ as $({\vec x}_2,r_2)$. This will be the
spacetime structure for which the holographic influence functional
is formulated.
\begin{figure}
\centering \scalebox{0.7}{\includegraphics{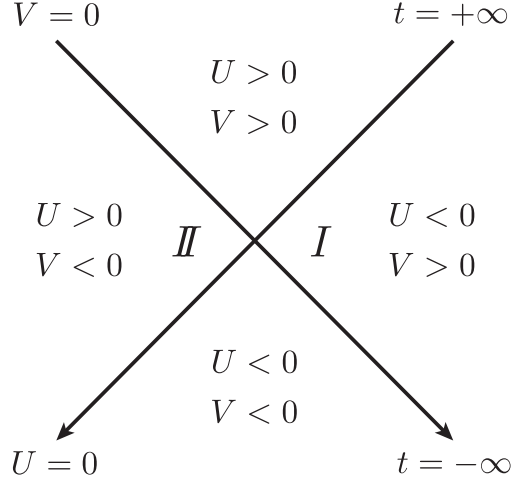}} \caption{The
Kruskal coordinates. The right quadrant corresponds to the field
variable labeled by $+$ and the left quadrant corresponds to the
field variable labeled by $-$.}
\end{figure}

\subsection{The semi-holographic approach for influence functional}
As in \cite{mirror}, the classical gravity action in~(\ref{gravity
action}) is  the linearized Nambu-Goto action for the probed
string in the background (\ref{lifshitz bh}), and is given by \be
  S_{NG}\approx - \frac1{4\pi\alpha'}\int dr \, dt \,
\bigg( r^{z+3} f(r) \partial_rX^I \partial_rX'^I-
\frac{\partial_tX^I\partial_tX^I}{f(r){r^{z-1}}}\bigg) \, ,
  \label{NG}
  \ee
where $X^I(t,r)$ is the linearized string perturbations in the
static gauge. Since the motion of string in different directions
decoupled, without losing generality we consider its motion along
one of the directions denoted by $X(t,r)$. The equation of motion
of the string in the frequency space,
$X(t,r)=X_{\omega}(r)e^{-i\omega t}$, is given by
 \be
  \label{NG with T}
 \frac{\partial}{\partial r}\biggl[ r^{z+3}f(r)\frac{\partial }{\partial
 r} X_{\omega}(r) \biggr]+\frac{\omega^2}{r^{z-1}f(r)} X_{\omega}(r)=0 \,
 .
  \ee
Near the horizon, the solution behaves like $X_{\omega}(r)\propto
e^{\pm i\omega r_*}$. Thus  two linearly independent solutions,
$\mathcal{X}_{\omega}(r)$ and $\mathcal{X}_{\omega}^*(r)$ can be found with the properties
$\mathcal{X}_{\omega}(r)_{\substack{
   \propto \\
   r\rightarrow r_h
  }} e^{+i\omega r_*}$ and
$\mathcal{X}^*_{\omega}(r)_{\substack{
   \propto \\
   r\rightarrow r_h
  }} e^{-i\omega r_*}$. We then choose a normalization so that $\mathcal{X}_{\omega}(r_b)=1$.
As in \cite{Son_09},  now the general solution in the extended black
hole can be parametrized as
    \bea
    &&Q^+(\omega,r_1)=a(\omega)\mathcal{X}_{\omega}(r_1)+b(\omega)\mathcal{X}_{\omega}^*(r_1)\, ,\nonumber\\
    &&Q^-(\omega,r_2)=a(\omega)\alpha_{\omega}\mathcal{X}_{\omega}(r_2)+b(\omega)\beta_{\omega}\mathcal{X}_{\omega}^*(r_2)
  \, .  \eea
Four boundary conditions are needed  to fix all the parameters. Two
of them are imposed at $r=r_b$
   \be
   \label{bcrb}
   Q^+(\omega,r_b)=q^+(\omega)\, ,\,\,\, Q^-(\omega,r_b)=q^-(\omega)  \, ,
   \ee
which are interpreted as the boundary sources and identified as
the position of the Brownian particle.  Thus we find
   \bea
   &&a(\omega)=\frac1{\alpha_{\omega}-\beta_{\omega}}q^-(\omega)-\frac{\beta_{\omega}}{\alpha_{\omega}-\beta_{\omega}}q^+(\omega)\, ,\nonumber\\
   &&b(\omega)=\frac{\alpha_{\omega}}{\alpha_{\omega}-\beta_{\omega}}q^+(\omega)-\frac1{\alpha_{\omega}-\beta_{\omega}}q^-(\omega)
  \, , \eea
with  $\alpha_{\omega}$ and $\beta_{\omega}$ undetermined. This
general solution is then plugged into the classical action in
(\ref{NG}),  resulting in a boundary term at $r=r_b$ as
\begin{align}\label{NGos}
     S_{NG}&=-\frac{r_b^{z+3}}{4\pi\alpha'}\int \frac{d\omega}{2\pi}\Bigl[Q^+(-\omega,r_b)\partial_r Q^+(\omega,r_b)-Q^-(-\omega,r_b)\partial_r Q^-(\omega,r_b)\Bigr]\nonumber\\
     &=-\frac{1}{2}\int\frac{d\omega}{2\pi}\biggl\{q^+(-\omega)\Bigl[A_{\omega}\operatorname{Re}G_R(\omega)+iB_{\omega}\operatorname{Im}G_R(\omega)\Bigr]q^+(\omega)\biggr.\nonumber\\
     &\qquad\qquad\quad+q^-(-\omega)\Bigl[C_{\omega}\operatorname{Re}G_R(\omega)+iD_{\omega}\operatorname{Im}G_R(\omega)\Bigr]q^-(\omega)\nonumber\\
     &\qquad\qquad\quad-q^+(-\omega)\Bigl[E_{\omega}\operatorname{Re}G_R(\omega)+iF_{\omega}\operatorname{Im}G_R(\omega)\Bigr]q^-(\omega)\nonumber\\
     &\qquad\qquad\quad-\biggl.q^-(-\omega)\Bigl[G_{\omega}\operatorname{Re}G_R(\omega)+iH_{\omega}\operatorname{Im}G_R(\omega)\Bigr]q^+(\omega)\biggr\}
     \, ,
\end{align}
where we have defined $G_R(\omega)=\frac{r_b^{z+3}}{2\pi\alpha'}
\mathcal{X}_{-\omega}(r_b)\partial_r\mathcal{X}_{\omega}(r_b)$, and
this coincides with the retarded Green function constructed by the
prescription in \cite{Son_09}. The forms of the parameters
$A_{\omega}$, $B_{\omega}$, ..., $H_{\omega}$ in term of
$\alpha_{\omega}$ and $\beta_{\omega}$ are given in Appendix
\ref{appen1}. We can identify the action in (\ref{NGos}) as the
influence functional ${\mathcal{F}}(q^+,q^-)$ with $q^+$ and $q^-$
being as sources in two respective branches of the closed-time-path
contour. The field correlators are obtained by taking functional
derivative
  \be
  \label{SK}
  G^{a b}(\omega)=\frac{\delta}{\delta q^{a}}\frac{\delta}{\delta
  q^{b}}S_{NG}[q^+,q^-] \,
  \ee
where $a$ and $b$ label the branches $\pm$.

The idea now is to propose a prescription to explicitly write down
the holographic influence functional \eqref{NGos}, that is, to
determine $\alpha_{\omega}$ and $\beta_{\omega}$. The rationale of
the following prescription is directly using basic properties of
the influence functional to set up the dictionary. For the thermal
states of the bosonic field the influence functional obeys the
unitarity and periodicity conditions as in
Eqs.~(\ref{periodicity}) and (\ref{u-identity}). We first impose
the unitarity condition encoded in (\ref{u-identity}) and find
  \bea
  &&A_{\omega}+C_{\omega}+E_{\omega}+F_{\omega}=0 \, ,\nonumber\\
  &&B_{\omega}+E_{\omega}+F_{\omega}+G_{\omega}=0\, .
  \eea
The solution is either $\alpha_{\omega}=1$,
$\beta_{-\omega}=\beta^{-1}_{\omega}$ or $\beta_{\omega}=1$,
$\alpha_{-\omega}=\alpha^{-1}_{\omega}$ where both $\alpha_{\omega}$
and $\beta_{\omega}$ are real functions of $\omega$ (see Appendix
\ref{appen1}). These two solutions will give the same influence
functional as we will see later. Let us choose the first one and we
have
  \bea
  \label{KMS0}
  G^{++}(\omega)&=&\operatorname{Re}G_R(\omega)+\frac{\beta_{\omega}+1}{\beta_{\omega}-1}\,i\operatorname{Im}G_R(\omega)\, ,\nonumber\\
  G^{--}(\omega)&=&-\operatorname{Re}G_R(\omega)+\frac{\beta_{\omega}+1}{\beta_{\omega}-1}\,i\operatorname{Im}G_R(\omega)\, , \nonumber\\
  G^{+-}(\omega)&=&\frac{2\beta_{\omega}}{\beta_{\omega}-1}\,i\operatorname{Im}G_R(\omega)\, ,\nonumber\\
  G^{-+}(\omega)&=&\frac{2}{\beta_{\omega}-1}\,i\operatorname{Im}G_R(\omega)
  \,.
  \eea
Here $\beta_{\omega}$ remains undetermined. Finally since the
periodicity condition in (\ref{periodicity}) requires
  \be\label{erdsfd}
  \frac{G^{+-}(\omega)}{G^{-+}(\omega)}=e^{-\frac{\omega}{T}}\,,
  \ee
we use it to fix $\beta_{\omega}$ and find
$\beta_{\omega}=e^{-\frac{\omega}{T}}$. As for the general
contour~\cite{semenoff} with a finite $\sigma$ seen in Fig.~1, the
unitarity condition becomes
  \be
  G^{++}(\omega)-e^{\omega\sigma}G^{+-}(\omega)-e^{-\omega\sigma}G^{-+}(\omega)+G^{--}(\omega)=0
  \, .
  \ee
Accordingly the periodicity condition for the thermal bath is
modified to the form
  \be
  \frac{G^{+-}(\omega)}{G^{-+}(\omega)}=e^{-\frac{\omega}{T}+2\sigma\omega}
  \, .
  \ee
Putting all together, we will arrive at
$\alpha_{\omega}=e^{-\omega\sigma}$ and
$\beta_{\omega}=e^{-\omega\sigma}e^{\frac{\omega}{T}}$. The
corresponding nonequilibrium correlators for general $\sigma$ is
obtained as
  \bea
  \label{SKs}
  G^{++}(\omega)&=&\operatorname{Re}G_R(\omega)+(1+2n)\,i\operatorname{Im}G_R(\omega)\, ,\nonumber\\
  G^{--}(\omega)&=&-\operatorname{Re}G_R(\omega)+(1+2n)\,i\operatorname{Im}G_R(\omega)\, , \nonumber\\
  G^{+-}(\omega)&=&-2n\,e^{\omega\sigma}\,i\operatorname{Im}G_R(\omega)\, ,\nonumber\\
  G^{-+}(\omega)&=&-2(1+n)\,e^{-\omega\sigma}\,i\operatorname{Im}G_R(\omega)
  \, ,
  \eea
where $n=(e^{\frac{\omega}{T}}-1)^{-1}$ is the occupation number.
Especially, in the context of quantum decoherence, two particular
combinations of these correlators are useful,
\begin{align}
    G_{R} (t-t')&=\bigg\{ G^{++} ( t-t') -G^{+-} (t-t') \bigg\}\,,\\
    G_{H} (t-t')&=\frac{i}{4} \bigg\{ G^{++} ( t-t') +G^{+-} (t-t') + G^{++} ( t-t') +G^{+-} (t-t') \bigg\}\,.
\end{align}
Therefore the holographic influence functional is fully
constructed by the prescription we propose here.

\section{quantum decoherence induced by strongly coupled fields}\label{sec3}
 We now apply the obtained holographic
influence functional to the quantum decoherence problem, in
particular, for the Brownian particle moving in the strongly coupled
quantum critical theories. For the linear coupling between the
system and environment, the reduced density matrix and the influence
functional are found  in (\ref{evolveelectron}) and (\ref{inf})
respectively.  We then follow the approach in \cite{Lee_06} to  set
up our interference experiment.  To proceed, we express the
influence functional  in a form  of its phase and modulus by:
\begin{equation}
    \mathcal{F}[\,q^+,q^-]=\exp\Big\{\mathcal{W}[\,q^+,q^-]+i\,\Phi[\,q^+,q^-]\Bigr\}\,,
\end{equation}
where the real functionals $\mathcal{W}$ and $\Phi$ are
\begin{align}
    \Phi[\,q^+,q^-]&=\frac{1}{2}\int\!dt \!\!\int\!dt'\Bigl[\,q^+(t)-q^-(t')\Bigr]G_{R}(t-t')\Bigl[\,q^+(t) + q^-(t')\Bigr]\,,\label{phase}\\
    \mathcal{W}[\,q^+,q^-]&=-\frac{1}{2}\int\!dt \!\!\int\!dt'\Bigl[\,q^+(t)-q^-(t')\Bigr]G_{H}(t-t')\Bigl[\,q^+(t) - q^-(t')\Bigr]\,.\label{decoherence}
\end{align}
The retarded Green's function and the Hadamard function of
the environmental field are respectively defined by
\begin{eqnarray}
    G_{R} (t-t')&=&-i\,\theta(t-t')\,\bigl<\left[{{F}} (t),{{F}} (t')\right]\bigr>\,,\label{commutator}\\
    G_{H} (t-t')&=& \frac{1}{2}\,\bigl<\left\{{{F}} (t),{{F}} (t')\right\}\big>\,.
    \label{anticommutator}
\end{eqnarray}
For a given initial state for the particle, the reduced density
matrix at time $t_f$ can be obtained from~\eqref{evolveelectron} when the path integration over
${q}^{\pm}$ in~\eqref{propagator} is carried out. Expressed explicitly in terms of the phase and modulus of the
influence functional, the reduced density matrix now becomes
\begin{eqnarray}
    \rho_r({q}_f,\tilde{{q}}_f,t_f)&=&\int\!d{q}_1\,d{q}_2\left[\int^{{q}_f}_{{q}_1}\!\!\mathcal{D}{q}^+\!\!\int^{\tilde{{q}}_f}_{{q}_2}\!\!\mathcal{D}{q}^-\;\exp\biggl\{i\int_{t_i}^{t_f}dt\left(L_{q}[{q}^+]-L_{q}[{q}^-]\right)\biggr\}\right.\nonumber\\
            &&\qquad\qquad\left.\,\times\,\exp\biggl\{\mathcal{W}[{q}^+,{q}^-]\biggr\}\exp\biggl\{i\,\Phi[{q}^+,{q}^-]\biggr\}\right]\rho_{q}({q}_1,{q}_2,t_i)\,.\label{evolve}
\end{eqnarray}
Suppose that the initial  state $\bigl|\Psi(t_i)\bigr>$
of the particle is a coherent superposition of two localized
states, which will travel along the two non-overlapping paths ${C}_1$
and ${C}_2$. If both states leave the same spatial point at the moment
$t_i$,
\begin{equation}
    \bigl|\Psi(t_i)\bigr>=\bigl|\psi_1(t_i)\bigr>+\bigl|\psi_2(t_i)\bigr>\,,
\end{equation}
then the initial density matrix of the state can be written as
\begin{eqnarray}
    \rho_q(t_i)=\bigl|\Psi(t_i)\bigr>\bigl<\Psi(t_i)\bigr|=\rho_{11}(t_i)+\rho_{22}(t_i)+\rho_{21}(t_i)+\rho_{12}(t_i)\,,
\end{eqnarray}
where
$\rho_{mn}(t_i)=\bigl|\psi_m(t_i)\bigr>\bigl<\psi_n(t_i)\bigr|$. The
term $\rho_{21}+\rho_{12}$ accounts for quantum interference,
because when the density matrix is realized in the coordinate basis,
we have
\begin{equation}
    \bigl<{q}_i\big|\rho_q(t_i)\big|{q}_i\bigr>=\left|\psi_1({q}_i,t_i)\right|^2+\left|\psi_2({q}_i,t_i)\right|^2+2\,\operatorname{Re}\left\{\psi_2^*({q}_i,t_i)\psi_1^{\vphantom{*}}({q}_i,t_i)\right\}\,,
\end{equation}
which expresses the probability of finding the particle at
$(t_i,{q}_i)$ in the superposed state.

Next let these two localized states $\bigl|\psi_m\bigr>$ move along
the respective path $C_{m}$, and then they are recombined at the location
${q}_f$ at later time $t_f$. The corresponding
spacetime paths are drawn in Fig.~3. The interference pattern of
the superposed state at $t_{f}$ is given by the diagonal
elements of the reduced density matrix
$\bigl<{q}_f\big|\rho_r(t_f)\big|{q}_f\bigr>=\rho_r({q}_f,{q}_f,t_f)$\,.
\begin{figure}
\centering
    \scalebox{0.5}{\includegraphics{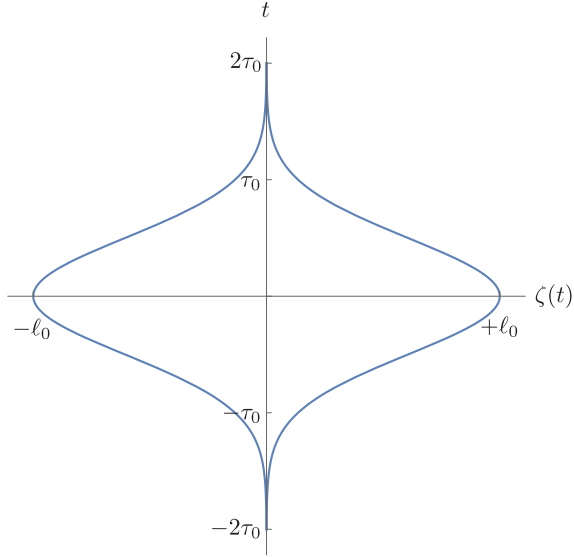}}
\caption{The spacetime paths of the wavepackets for an
interference experiment.}
\end{figure}
The mean trajectory of the wave packets $\bigl|\psi_m\bigr>$ follows
the classical path constrained by an appropriate external potential.
We assume that the finite spread of the wave packet of the state can
be legitimately neglected as long as the de Broglie wavelength,
$\lambda_{\mathrm{dB}}$ is much shorter than the characteristic
length scale associated with the accuracy of the measurement
$\ell_0$, about the same order as travel distance for the states to
recombine. Thus, when $\ell_0 \gg\lambda_{\mathrm{dB}}$, the wave
packet can be seen to remain quite sharply peaked in its position
and momentum, thus we may ignore the effect of wavefunction
spreading.
 As such, the leading effect of the decoherence can be obtained by evaluating the propagating
function \eqref{propagator} along a prescribed classical path. The
diagonal components of the reduced density matrix
$\rho_r({q}_f,{q}_f,t_f)$ now becomes
\begin{equation}
    \rho_r({q}_f,{q}_f,t_f)=\bigl|\psi_1({q}_f,t_f)\bigr|^2+\bigl|\psi_2({q}_f,t_f)\bigr|^2+2\,e^{\mathcal{W}[\,\bar{q}^+,\bar{q}^-]}\;\operatorname{Re}\left\{e^{i\,\Phi[\,\bar{q}^+,\bar{q}^-]}\psi_1^{\vphantom{*}}({q}_f,t_f)\,\psi_{2}^{*}({q}_f,t_f)\right\}\,,
\end{equation}
where the $\mathcal{W}$ and $\Phi$ functionals are evaluated along
the classical trajectories, ${C}_1=\bar{q}^{+}$ and
${C}_2=\bar{q}^{-}$.

The decoherence functional $\mathcal{W}$, determined by the
expectation value of the anti-commutator of environmental field
operators, reveals decoherence between particle states,
while its phase functional $\Phi$, related to  the expectation value
of the commutator of field operators, results in an overall phase
shift in the interference pattern. Both effects arise from the
interaction with quantum fields. It can be seen that the effects of
quantum decoherence and the phase shift are likely to be related by
the fluctuation-dissipation relation
  \be
  G_H(\omega)=- \coth \bigg[\frac{\omega}{ 2T}\bigg] \operatorname{Im}G_R(\omega)
  \,. \label{FDR_T}
  \ee
At zero temperature $T\to0$, the relation reduces to
\be
  G_H(\omega)=  -\operatorname{sgn}(\omega) \operatorname{Im}G_R(\omega) \, .
  \ee
Here we consider the prescribed trajectories $\bar{q}^{\pm}= \pm
\zeta (t) $ of the localized states. The path function $\zeta(t)$
is required to be sufficiently smooth and is chosen to take the
form,
\begin{equation}\label{path}
    \zeta(t)={ \ell}_0\,e^{-\frac{t^2}{\tau_0^2}}\,,
\end{equation}
where $2\ell_0$ is the effective path separation and $2\tau_0$ is
the effective flight time as in \cite{Lee_06} (See Fig.~3). We then
have the decoherence functional from~(\ref{decoherence})
   \be \label{W}
   \mathcal{W}=- \int\frac{d\omega}{2\pi}\; G_H(\omega) \int
   dt \,\bigg\vert \, \zeta(t) \, e^{-i\omega t} \bigg\vert^2 <0\, .
   \ee
Since these two prescribed classical trajectories are symmetric
with respect to the initial position, it implies
$q^{+}(t)+q^{-}(t)=0$ and in turns we will not see any phase
shift, that is, $\Phi =0$ from~(\ref{phase}). We now use the
holographic influence functional to evaluate $\mathcal{W}$.

\subsection{Holography at zero temperature}
In the general Liftshitz black hole background, the analytic
expression for $G_R(\omega)=\frac{r_b^{z+3}}{2\pi\alpha'}\,
{X}_{-\omega}(r_b)\partial_r {X}_{\omega}(r_b)$ cannot be obtained
except in the zero temperature limit where it is given by (see
\cite{mirror} and \cite{Tong_12})
   \be
   G_R(\omega)=\frac{\omega r_b^2}{2\pi\alpha'}\frac{H^{(1)}_{\frac1z-\frac12}(\frac{\omega}{zr_b^z})}{H^{(1)}_{\frac1z+\frac12}(\frac{\omega}{zr_b^z})}
\, .
   \ee
Here $H^{(1)}_{\nu}(z)$ is Hankel's function and it takes values
in the principle branch $-\pi<\arg(z)<\pi$ in order for
$G_R(\omega)$ to be analytic on the upper half of $\omega$-plane.
The Hadamard function can be found from the retarded Green's
function by means of the fluctuation-dissipation relation. Its
leading contribution in the small frequency limit is
  \be
  G_H(\omega)\simeq
  \frac1{2\pi\alpha'}\, \frac{\cos(\frac{\pi}{z})}{(2z)^{\frac{z}{2}}}\, \frac{\Gamma(\frac12-\frac1z)}{\Gamma(\frac12+\frac1z)}\, \operatorname{sgn}(\omega)\,
  \omega^{1+\frac2z} + {\mathcal O}(\omega^2/r_b^{2z}).
  \ee
For the general $z$ the leading effect to the decoherence functional
is found to be
    \be
\mathcal{W}\simeq-\frac{2}{\pi\alpha'}\,
\frac{\cos(\frac{\pi}{z})}{(z)^{\frac{z}{2}}}\,
\frac{\Gamma(1+\frac1z)
\Gamma(\frac12-\frac1z)}{\Gamma(\frac12+\frac1z) } \,
\frac{\ell_0^2}{\tau_0^{\frac2z}}+ {\mathcal O}(1/r_b^{2z}\tau_0^2) \,.
    \ee
The $\mathcal{W}$ functional is smooth cross $z=2$. In \cite{mirror}
the holographic method is used to study fluctuations and dissipation
of an $n$-dimensional moving-mirror coupled to quantum critical
theories. The effects of the environment on the system are
classified according to the low-frequency  behavior of the the
friction term $\gamma \omega^{k+1}$, as subohmic, ohmic, and
supraohmic for $ k<0$, $k=0$ and $k>0$ respectively. We then found
that the low-frequency expansion of the damping term by the
holographic approach gives $k=({n+2})/{z}$. For a
point particle, which corresponds to the $n=0$ case, the dissipative dynamics due
to these quantum critical theories with $z$ not much larger than $2$
is supraohmic. However, when $z\gg 2$,
the dissipative dynamics then demonstrates the ohmic behavior. Thus, for
a fixed travel time $\tau_0$, the decoherence effect
 decreases as  $z$ increases toward the ohmic situation. This is one
 of our main results.

In particular, for $z=1$ (the relativistic environmental field), the
decoherence function reduces to
  \be \mathcal{W} \simeq
-\frac{4}{\pi\alpha'}\frac{\ell_0^2}{\tau_0^{2}} + {\mathcal
O}(1/r_b^2\tau_0^2)\,
  \ee
with velocity squared dependence.  This is the same as the result
in~\cite{Lee_05}, but the proportionality constant are different
between the strongly coupled environment and the free field
background, as expected. Notice that it will always show enhancement in the
strong field theory with coupling
 $\lambda>>1$ through the relation
$\lambda=L^4/\alpha'^2$, as the consequence of  AdS/CFT correspondence, where
$L$ is the curvature radius in the Lifshitz background.

\subsection{Holography in high temperature}
In the finite temperature case, the corresponding retarded
Green's function (the response function) has been studied in
\cite{Tong_12,mirror}, and in the low frequency and high
temperature approximation
($\frac{r_h}{r_b}>>\frac{\omega}{r_b^z}$), this is given by
 \be
  G^{(T)}_{R}(\omega)=-\gamma_{ T}(z)\, i\,\omega+m_{T}(z) (i\omega)^2+\mathcal{O}(\omega^3)\,,
  \ee
 where
  \be \label{m_gamma_T} m_{
  T}(z)=\frac{1}{2\pi\alpha'}\frac{1}{r_b^{z-2}}
  \bigg(\frac{r_h}{r_b}\bigg)^{4}\Bigl[(2+z)-\kappa
  r_b^{z+2}\Bigr]\, , \qquad\qquad \gamma_{ T}(z)=
  \frac1{2\pi\alpha'}r_h^{2} \,. \ee
The parameter $\kappa$ is the integration constant depends on the
detail of black hole geometry. The inertial mass $m_{ T}$ and the
damping coefficient $\gamma_{T}$ have the temperature dependence
through the black hole temperature~(\ref{BHT}). Since the damping
term has linear $\omega$ dependence, the stochastic dynamics of the
particle in the thermal environment will be ohmic. By the
fluctuation-dissipation relation (\ref{FDR_T}) the Hadamard function
at high $T$ is obtained as
    \be
    G^{(T)}_H(\omega)\simeq
    2T\gamma_T(z)=\frac{2T}{2\pi\alpha'}\left(\frac{4\pi
    T}{z+3}\right)^{\frac2z} \, .
    \ee
This gives
    \be
    \mathcal{W}_T\simeq - \frac{16\pi
    \ell_0^2\tau_0}{\alpha'}\sqrt{\frac{\pi}{2}}T\left(\frac{4\pi
    T}{z+3}\right)^{\frac2z}\, .
    \ee
Compared with the case of the free environment field, the same
fluctuation-dissipation relation at finite
temperature~(\ref{FDR_T}) is found in \cite{Zurek_97, Lee_08}.
However, a distinctive feature of the free field case is that the
linear coupling between the particle and the environment field
leads to a damping coefficient $\gamma$ that does not depend on
the state of the environment and thus is independent of
temperature~\cite{Zurek_97}. Thus, in the free field case, the
$\mathcal{W}$ functional reveals linear dependence in temperature.
This is in dramatic difference from the strongly coupled
environment where Eq.~\eqref{m_gamma_T} shows the temperature
dependence of $\gamma$. The additional $T$ dependence certainly
arises from the effects from strong self-interaction of the
environment field, and  can not be obtained by a weak coupling
expansion. Therefore, the decoherence effect will be enhanced from
not only the strong coupling of the environment field itself, but
also the finite temperature effects. Similar to the zero
temperature case, the decoherence effect also decreases as $z$
increases. In order to reduce the decoherence effects from quantum
critical theories, the larger value of the dynamical exponent $z$
is suggested.

\section{concluding remarks}\label{sec4}
Understanding  how the  environment affects the system of interest
is the main concerns in nonequilibrium statistical mechanics. By
tracing out the environment degrees of freedom, its effects on the
system can be encoded in the influence functional. In this paper a
probed particle in the strongly coupled quantum critical theories
with general dynamical scaling $z$ is considered. The holographic
setup is a probed string in the classical Lifshitz geometry. We
first adapt the approach in \cite{Son_09} to construct the
holographic influence functional.  However it is modified based on
the field-theoretic consideration. We then apply the holographic
influence functional to study the decoherence of the particle
induced by strongly coupled quantum critical theories. We find that
the decoherence effect gets enhanced  as the coupling of the
environmental field becomes strong. Nevertheless, when the dynamical
scaling $z$ increases, the quantum critical theories transit from a
supraohmic to an ohmic environment, and the decoherence effect
becomes less significant.  Additionally the
decoherence effect will be enhanced not only by its strong coupling
constant but also from the finite temperature effects that certainly
can not be achieved by a weak coupling expansion.  Similar to the
zero temperature case, the larger value of the dynamical exponent
$z$ is suggested to reduce the decoherence effect at finite
temperature from quantum critical theories.

Furthermore, this revised prescription may pave a
way for extending the holographic method of influence functional to
more general states such as squeezed vacuum state~\cite{squeeze} so
as to consider the potential ``subvacuum" phenomena~\cite{Lee_12}.
Manipulating the quantum field may give rise to suppression of its
vacuum uncertainties. One of the known examples is the squeezed
vacuum state, from which  the decoherence of the particle can be
suppressed below the level due to the pure vacuum state, leading to
the so-called subvacuum phenomena~\cite{HS}. It is also known that
subvacuum phenomena can not last for an arbitrarily long period of
time. However, so far all the previous studies  consider only the  free field,
so its effect on the system is expected to be weak. Thus, our future work will explore this phenomenon in a strongly coupled field
theory in the hope that it will  be significantly amplified so as to be more observable in the experiment realization.

\begin{acknowledgments}
This work was supported in part by the National Science Council,
Taiwan.
\end{acknowledgments}
\appendix
\section{coefficients for the holographic influence
functional}\label{appen1} In this appendix, we write down the
Schwinger-Keldysh correlators in term of two unknown parameters
$\alpha_{\omega}$ and $\beta_{\omega}$. We will determine one of them
using the
  unitarity requirement in the zero $\sigma$ case. In the section III-B, the functions
  $A_{\omega}$, $B_{\omega}$, \ldots, $H_{\omega}$ are
   \bea
   A_{\omega}&=&\frac{(1-\alpha_{\omega}\alpha_{-\omega})\beta_{\omega}\beta_{-\omega}-(1-\alpha_{\omega}\beta_{-\omega})\beta_{\omega}\alpha_{-\omega}-(1-\beta_{\omega}\alpha_{-\omega})\alpha_{\omega}\beta_{-\omega}+(1-\beta_{\omega}\beta_{-\omega})\alpha_{\omega}\alpha_{-\omega}}{(\alpha_{\omega}-\beta_{\omega})(\alpha_{-\omega}-\beta_{-\omega})}\,,\nonumber\\
   B_{\omega}&=&\frac{(1-\alpha_{\omega}\alpha_{-\omega})\beta_{\omega}\beta_{-\omega}-(1-\alpha_{\omega}\beta_{-\omega})\beta_{\omega}\alpha_{-\omega}+(1-\beta_{\omega}\alpha_{-\omega})\alpha_{\omega}\beta_{-\omega}-(1-\beta_{\omega}\beta_{-\omega})\alpha_{\omega}\alpha_{-\omega}}{(\alpha_{\omega}-\beta_{\omega})(\alpha_{-\omega}-\beta_{-\omega})}\,,\nonumber\\
   C_{\omega}&=&\frac{-(1-\alpha_{\omega}\alpha_{-\omega})\beta_{\omega}+(1-\alpha_{\omega}\beta_{-\omega})\beta_{\omega}+(1-\beta_{\omega}\alpha_{-\omega})\alpha_{\omega}-(1-\beta_{\omega}\beta_{-\omega})\alpha_{\omega}}{(\alpha_{\omega}-\beta_{\omega})(\alpha_{-\omega}-\beta_{-\omega})}\,,\nonumber\\
   D_{\omega}&=&\frac{-(1-\alpha_{\omega}\alpha_{-\omega})\beta_{\omega}+(1-\alpha_{\omega}\beta_{-\omega})\beta_{\omega}-(1-\beta_{\omega}\alpha_{-\omega})\alpha_{\omega}+(1-\beta_{\omega}\beta_{-\omega})\alpha_{\omega}}{(\alpha_{\omega}-\beta_{\omega})(\alpha_{-\omega}-\beta_{-\omega})}\,,\nonumber\\
   E_{\omega}&=&\frac{-(1-\alpha_{\omega}\alpha_{-\omega})\beta_{-\omega}+(1-\alpha_{\omega}\beta_{-\omega})\alpha_{-\omega}+(1-\beta_{\omega}\alpha_{-\omega})\beta_{-\omega}-(1-\beta_{\omega}\beta_{-\omega})\alpha_{-\omega}}{(\alpha_{\omega}-\beta_{\omega})(\alpha_{-\omega}-\beta_{-\omega})}\,,\nonumber\\
   F_{\omega}&=&\frac{-(1-\alpha_{\omega}\alpha_{-\omega})\beta_{-\omega}+(1-\alpha_{\omega}\beta_{-\omega})\alpha_{-\omega}-(1-\beta_{\omega}\alpha_{-\omega})\beta_{-\omega}+(1-\beta_{\omega}\beta_{-\omega})\alpha_{-\omega}}{(\alpha_{\omega}-\beta_{\omega})(\alpha_{-\omega}-\beta_{-\omega})}\,,\nonumber\\
   G_{\omega}&=&\frac{(1-\alpha_{\omega}\alpha_{-\omega})-(1-\alpha_{\omega}\beta_{-\omega})\alpha_{-\omega}-(1-\beta_{\omega}\alpha_{-\omega})+(1-\beta_{\omega}\beta_{-\omega})}{(\alpha_{\omega}-\beta_{\omega})(\alpha_{-\omega}-\beta_{-\omega})}\,,\nonumber\\
   H_{\omega}&=&\frac{(1-\alpha_{\omega}\alpha_{-\omega})-(1-\alpha_{\omega}\beta_{-\omega})\alpha_{-\omega}+(1-\beta_{\omega}\alpha_{-\omega})-(1-\beta_{\omega}\beta_{-\omega})}{(\alpha_{\omega}-\beta_{\omega})(\alpha_{-\omega}-\beta_{-\omega})}\,.
   \eea
By  (\ref{SK}) we can write the Schwinger-Keldysh
correlators as
  \bea
  G^{++}(\omega)&=&A(\omega)\operatorname{Re}G_R(\omega)+i\,B(\omega)\operatorname{Im}G_R(\omega)\,,\nonumber\\
  G^{--}(\omega)&=&C(\omega)\operatorname{Re}G_R(\omega)+i \,D(\omega)\operatorname{Im}G_R(\omega)\,,\nonumber\\
  G^{+-}(\omega)&=&-E(\omega)\operatorname{Re}G_R(\omega)-i\, F(\omega)\operatorname{Im}G_R(\omega)\,,\nonumber\\
  G^{-+}(\omega)&=&-G(\omega)\operatorname{Re}G_R(\omega)-i\, H(\omega)\operatorname{Im}G_R(\omega)\,.
  \eea
 The unitarity condition for zero $\sigma$ says that
  \be
  \label{unitary}
  G^{++}(\omega)-G^{+-}(\omega)-G^{-+}(\omega)+G^{--}(\omega)=0\,.
  \ee
In the following we will assume that $\alpha_{\omega}$ and
$\beta_{\omega}$ are real. From the unitarity and the periodicity
conditions we will find that this is the consistent assumption.
Thus the real part of this equation gives
  \bea
  A_{\omega}+C_{\omega}+E_{\omega}+F_{\omega}&=&(1-\alpha_{\omega}\alpha_{-\omega})(\beta_{\omega}\beta_{-\omega}-\beta_{\omega}-\beta_{-\omega}+1)\nonumber\\
  &&+(1-\alpha_{\omega}\beta_{-\omega})(-\beta_{\omega}\alpha_{-\omega}+\beta_{\omega}+\alpha_{-\omega}-1)\nonumber\\
  &&+(1-\beta_{\omega}\alpha_{-\omega})(-\alpha_{\omega}\beta_{-\omega}+\alpha_{\omega}+\beta_{-\omega}-1)\nonumber\\
  &&+(1-\beta_{\omega}\beta_{-\omega})(\alpha_{\omega}\alpha_{-\omega}-\alpha_{\omega}-\alpha_{-\omega}+1)\nonumber\\
  &=&0\,.
  \eea
One solution to this equation is
$\alpha_{-\omega}=\alpha^{-1}_{\omega}$ and
$\beta_{-\omega}=\beta^{-1}_{\omega}$. The imaginary part of (\ref{unitary}), $
A_{\omega}+C_{\omega}+E_{\omega}+F_{\omega}=0$, then yields to
  \be
  (1-\frac{\alpha_{\omega}}{\beta_{\omega}})(-\frac{\beta_{\omega}}{\alpha_{\omega}}+\beta_{\omega}+\frac{1}{\alpha_{\omega}}-1)-(1-\frac{\beta_{\omega}}{\alpha_{\omega}})(-\frac{\alpha_{\omega}}{\beta_{\omega}}+\alpha_{\omega}+\frac{1}{\beta_{\omega}}-1)=0\,.
  \ee
Either $\alpha_{\omega}=1$ or $\beta_{\omega}=1$ is a solution.

\section{using the boundary condition at the horizon}\label{appen2}
In this appendix we determine the solution of the equation
  (\ref{NG with T}) by imposing the boundary condition at the horizon, proposed in
  \cite{Son_09}. As discussed in the main text, we have the two linearly-independent solutions. One is $e^{-i\omega
t}\mathcal{X}_{\omega}(r)_{\substack{
   \longrightarrow \\
   r\rightarrow r_h
  }} e^{-i\omega(t+r_*)}=e^{-i\,\frac{\omega}{2\pi T}\,\ln V}$ and the other is $e^{-i\omega
t}\mathcal{X}_{\omega}^*(r)_{\substack{
   \longrightarrow \\
   r\rightarrow r_h
  }} e^{i\omega(-t+r_*)}=e^{i\,\frac{\omega}{2\pi T}\,\ln (-U)}$,
  where $U$ and $V$ are Kruskal coordinates defined in (\ref{Kruskal})

The general solution in the extended black
  hole background is
    \bea
    &&Q^+(\omega,r_1)=a(\omega)\mathcal{X}_{\omega}(r_1)+b(\omega)\mathcal{X}_{\omega}^*(r_1)\,,\nonumber\\
    &&Q^-(\omega,r_2)=c(\omega)\mathcal{X}_{\omega}(r_2)+d(\omega)\mathcal{X}_{\omega}^*(r_2)\,.
    \eea
We impose two boundary conditions at $r_1=r_b$ and $r_2=r_b$ as in (\ref{bcrb}). The other two boundary conditions are
imposed on the horizon. As proposed in \cite{Son_09}, in order to
obtain the right correlators, they required the modes with positive
(negative) frequency are purely incoming (outgoing). (By
definition, in the region $I\!\!I$, the outgoing wave is coming in to
the horizon). The modes that satisfy these boundary conditions are
    \bea
    &&e^{-i\omega t}\mathcal{X}_{\omega}(r_1)+e^{-\frac{\omega}{2T}}e^{-i\omega t}\mathcal{X}_{\omega}(r_2)_{\substack{
   \longrightarrow \\
   r\rightarrow r_h
  }}e^{-\frac{i\omega}{2\pi T}\ln(V)}\,,
    \eea
which is incoming with positive frequency (analytic in lower half
$V$-plane), and
    \bea
    &&e^{-i\omega t}\mathcal{X}_{\omega}^*(r_1)+e^{\frac{\omega}{2T}}e^{-i\omega t}\mathcal{X}_{\omega}^*(r_2)_{\substack{
   \longrightarrow \\
   r\rightarrow r_h
  }}e^{\frac{i\omega}{2\pi T}\ln(-U)}\,,
    \eea
which is outgoing with negative frequency (analytic in upper half
$U$-plane). This amounts to requiring that the solutions in the region
$I\!\!I$ is a continuation of the ones in the region $I$,
    \bea
    \label{modes}
    &&Q^+(\omega,r_1)=a(\omega)\mathcal{X}_{\omega}(r_1)+b(\omega)\mathcal{X}_{\omega}^*(r_1)\,,\nonumber\\
    &&Q^-(\omega,r_2)=a(\omega)e^{-\frac{\omega}{2T}}\mathcal{X}_{\omega}(r_2)+b(\omega)e^{\frac{\omega}{2T}}\mathcal{X}_{\omega}^*(r_2)\,.
    \eea
Then $a(\omega)$ and $b(\omega)$ can be determined by the boundary
conditions at $r=r_b$. We choose a normalization so that
$X_{\omega}(r=r_b)=1$, which gives
    \bea
    &&a(\omega)=q^+(\omega)(1+n)-
    q^-(\omega)e^{\frac{\omega}{2T}}n\,,\\
    &&b(\omega)=
    q^-(\omega)e^{\frac{\omega}{2T}}n-
    q^+(\omega)n\,,
    \eea
where $n=(e^{\omega/T}-1)^{-1}$. Plugging the solutions \eqref{modes}
into the gravity action and keeping terms to the quadratic order, we
obtain
\begin{align}
     Z(q^+,q^-)&=S_{gravity}\notag\\
     &=-\frac{r_b^{z+3}}{4\pi\alpha'}\int \frac{d\omega}{2\pi}\left(Q^+(-\omega,r_b)\partial r Q^+(\omega,r_b)-Q^-(-\omega,r_b)\partial rQ^-(\omega,r_b)\right)\nonumber\\
     &=-\frac{1}{2\pi\alpha'}\int\frac{d\omega}{2\pi}\biggl\{q^+(-\omega)\biggl[i\operatorname{Re}G_R(\omega)-(1+2n)\operatorname{Im}G_R(\omega)\biggr]q^-(\omega)\biggr.\nonumber\\
     &\qquad\qquad\qquad+q^-(-\omega)\biggl[-i\operatorname{Re}G_R(\omega)-(1+2n)\operatorname{Im}G_R(\omega)\biggr]q^-(\omega)\nonumber\\
     &\qquad\qquad\qquad-q^+(-\omega)\biggl[-2n\,e^{\frac{\omega}{2T}}\operatorname{Im}G_R(\omega)\biggr]q^-(\omega)\nonumber\\
     &\qquad\qquad\qquad-\biggl.q^-(-\omega)\biggl[-2(1+n)\,e^{-\frac{\omega}{2T}}\operatorname{Im}G_R(\omega)\biggr]q^+(\omega)\biggr\}\,,
\end{align}
where
$G_R(\omega)=\mathcal{X}_{\omega}(r_b)\partial_r\mathcal{X}_{\omega}(r_b)$
is the retarded correlator. This prescription naturally gives the
Schwinger-Keldysh correlators in the case $\sigma=\frac{1}{2T}$ as
compared with (\ref{SKs}).

\end{document}